\documentclass[a4paper,aps,prstper,amsmath,onecolumn,amssymb,nopacs,10pt]{revtex4-2}

\usepackage[margin=1in]{geometry}
\usepackage{booktabs}
\usepackage{siunitx}
\usepackage{graphicx}
\usepackage{amsmath, amssymb}
\usepackage{caption}
\usepackage{placeins}
\usepackage{hyperref}
\usepackage{subcaption}
\usepackage{booktabs}
\usepackage{threeparttable}

\graphicspath{{figures/}}
\begin{document}
\title{Active Learning vs Traditional Lecturing in Introductory Mechanics: A Pooled Pass-Rate Benchmark Under Common Departmental Assessments from a Latin American Institutional Change Initiative}
\author{Isaac Pérez Castillo}
\affiliation{Departamento de F\'isica, Universidad Autónoma Metropolitana-Iztapalapa, San Rafael Atlixco 186, Ciudad de México 09340, Mexico}
\author{Lidia Jiménez Lara}
\author{Orlando Guzman}
\author{Juan Ernesto Chavero Amador}
\author{Juan Miguel Carrillo Gil}
\author{Luis Alberto González Flores}
\affiliation{Departamento de F\'isica, Universidad Autónoma Metropolitana-Iztapalapa, San Rafael Atlixco 186, Ciudad de México 09340, Mexico}
\author{Miguel Angel Morales Olvera}
\author{Oscar Enrique Bonfil Urbalejo}
\affiliation{Coordinación Divisional de Docencia y Atención al Alumnado de la División de Ciencias Básicas e Ingeniería, Universidad Autónoma Metropolitana-Iztapalapa, San Rafael Atlixco 186, Ciudad de México 09340, Mexico}
\author{Eric Burkholder}
\affiliation{Department of Physics, Auburn University, Auburn, Alabama 36849, USA}

\begin{abstract}
Improving student success in introductory physics remains a persistent challenge despite the substantial progress made using research-based instructional practices in recent decades. Further, there is a dearth of literature which describes the implementation of these practices in the Latin American context, where resources for instructional change and implementation are often constrained. This study reports a transparent benchmark study of student passing outcomes in \textit{Elementary Mechanics I} at a large, public university in M\'exico between sections using Active Learning (AL) and sections using Traditional Lecturing (TL). The labels AL and TL are operational: they refer to section-level implementations by individual faculty rather than to perfectly standardized protocols, where instructors were given the agency to use a variety of techniques in both AL and TL. Using aggregated counts from group reports prepared by the course's coordinator and common departmental assessments--i.e., assessments written by a committee independent of the mode of instruction--we estimated pooled student-level pass probabilities for the first midterm exam, the second midterm exam, the global exam, and the final mark. Modality differences are summarized primarily by the risk difference, defined as $RD_a=p_{\mathrm{AL},a}-p_{\mathrm{TL},a}$ (given in percentage points), complemented by uncertainty quantification using (i) Wilson confidence intervals for pooled pass rates and (ii) a Bayesian reference analysis with Jeffreys priors for binomial proportions. Across assessments, pooled pass rates were higher under AL than under TL, with the strongest separation observed for the global (final comprehensive) exam and for the final mark. The $95\%$ confidence intervals for the global exam and the final mark did not include zero, even when using a random-intercept Bayesian model. We emphasize a constrained interpretation of these results consistent with the available data: the results provide a student-weighted benchmark of ``AL as implemented'' versus ``TL as implemented'' in this setting, but do not isolate the causal effect of any single instructional technique. Implications are discussed for departmental decision-making and for feasible next steps in evaluation, including improved student data collection in future terms and more robust qualitative analysis of the change initiative.
\end{abstract}

\maketitle

%%%%%%%%%%%%%%%%%%%%%%%%%%%%%%%%%%%%%%%%%%%%%%%%%%%%%%%%%%%%%%%%%%%%%%%%%%%%%%
\section{Introduction}
\label{sec:intro}
%%%%%%%%%%%%%%%%%%%%%%%%%%%%%%%%%%%%%%%%%%%%%%%%%%%%%%%%%%%%%%%%%%%%%%%%%%%%%%
Introductory physics courses occupy a consequential role in STEM curricula: they are both foundational for subsequent coursework and, for many students, the first sustained encounter with formal quantitative modeling. In many departments, however, these courses remain characterized by low pass rates and high repetition rates, which carry substantial academic and institutional costs. These challenges are often amplified in settings where entering preparation is heterogeneous \cite{PhysRevPhysEducRes.18.020124,burkholder2021mixed}, class sizes are large, and instructional resources are constrained---conditions that are common across a wide range of institutions, including many in Latin America. In such contexts, improving passing outcomes is not merely an administrative target; it is tightly linked to persistence in STEM pathways \cite{seymour2019talking}, equity of opportunity, and the feasibility of scaling high-quality instruction.

Over the past five decades, Physics Education Research (PER) and the broader science, technology, engineering, and mathematics (STEM) education literature have converged on a robust empirical finding: instructional approaches that engage students actively in sense-making, and are based in theories of cognition, tend to improve learning and reduce failure rates in large-enrollment courses relative to purely didactic lecturing \cite{Freeman2014,Theobald2020,Deslauriers2011,Hake1998,Mazur1997}. The umbrella term \emph{active learning} covers a family of practices rooted in constructivist theories \cite{piaget1977role} such as Ericsson's theory of ``deliberate practice'' \cite{ericsson1993role,PhysRevSTPER.11.020108}. These practices are broadly characterized by a shift from students listening to doing in the classroom, with the instructor orchestrating frequent formative assessment and targeted explanation. This can be through the use of in-class ``clicker'' questions, as in Peer Instruction, the use of educational simulations, and guided group problem-solving activities, to name a few examples. The literature also stresses that outcomes depend on local implementation: the specific practices adopted, the fidelity with which they are sustained, instructor experience, course constraints, and the degree to which assessments are aligned with intended learning goals \cite{WiemanPerkins2005}. These challenges can make it difficult for instructors to affect sustained instructional change \cite{dancy2010pedagogical}.

Though the evidence in support of active learning is overwhelming---to the point that many now argue that it should be treated as the standard of care when conducting research on learning strategies \cite{kramer2023establishing,wieman2014large}---there is simultaneously a dearth of research on the implementation of these practices in student populations beyond those at predominantly white, research-intensive, doctoral-granting universities in the United States and Europe \cite{kanim2020demographics}. This raises important questions about generalizability of the findings, particularly those concerning large-scale instructional change efforts. This paper documents the first results from an institutional-scale implementation of active learning at a large, urban research university in M\'exico which offers primarily STEM degrees, where improving gateway-course outcomes is an urgent priority. As a result, the academic unit has begun a coordinated instructional change effort for which introductory physics was selected as a testbed. We report results from the pilot term of this program \textit{Elementary Mechanics I} (EM1) in the Autumn trimester 2025 (coded at this institution as 25-O), using common departmental assessments that allow comparisons across sections under a shared grading standard. 

While the experimental design of comparing student outcomes in active learning classrooms to traditional lecturing is not novel, the context in which this study was conducted is comparatively understudied in the PER literature because of the student population and the context of the instructional change initiative.  The goal is to provide a transparent benchmark of student passing outcomes under \textit{AL as implemented} versus \textit{TL as implemented} in this setting, using official, aggregated records. Our research questions are: (1) Under a shared departmental assessment standard, how do pooled student-level pass probabilities differ between AL-labeled and TL-labeled sections across the major assessments of the course? (2) Are any observed differences consistent across midterm exams, the global exam, and the final mark? 

The remainder of the paper is organized as follows. Section~\ref{sec:initiative} describes the initiative and the operational meaning of the AL and TL labels in this implementation. Section~\ref{sec:data} documents the data provenance and caveats associated with aggregated reporting. Section~\ref{sec:analysis} presents the statistical analysis and discussion of the results, including both frequentist uncertainty intervals and a Bayesian reference analysis. Section~\ref{sec:conclusion} concludes with implications for instructional decision-making and priorities for future evaluation.

%%%%%%%%%%%%%%%%%%%%%%%%%%%%%%%%%%%%%%%%%%%%%%%%%%%%%%%%%%%%%%%%%%%%%%%%%%%%%%
\section{Background and Institutional Context}
\label{sec:initiative}
%%%%%%%%%%%%%%%%%%%%%%%%%%%%%%%%%%%%%%%%%%%%%%%%%%%%%%%%%%%%%%%%%%%%%%%%%%%%%%

At our institution, historical outcomes in the core physics curriculum have been a persistent concern for student progression and for the sustainability of teaching loads. In 2025, the Departamento de F\'isica at this university, under the lead of the new core curriculum coordinator, launched an active-learning initiative aimed at improving gateway-course outcomes while maintaining a common assessment standard. The implementation was designed as an institutional-scale intervention that includes instructor and teaching assistant (TA) preparation, explicit alignment of course activities with shared learning goals, and observational monitoring of classroom practices to document and support implementation across sections.

\begin{figure}[t]
\centering
\includegraphics[width=0.95\linewidth]{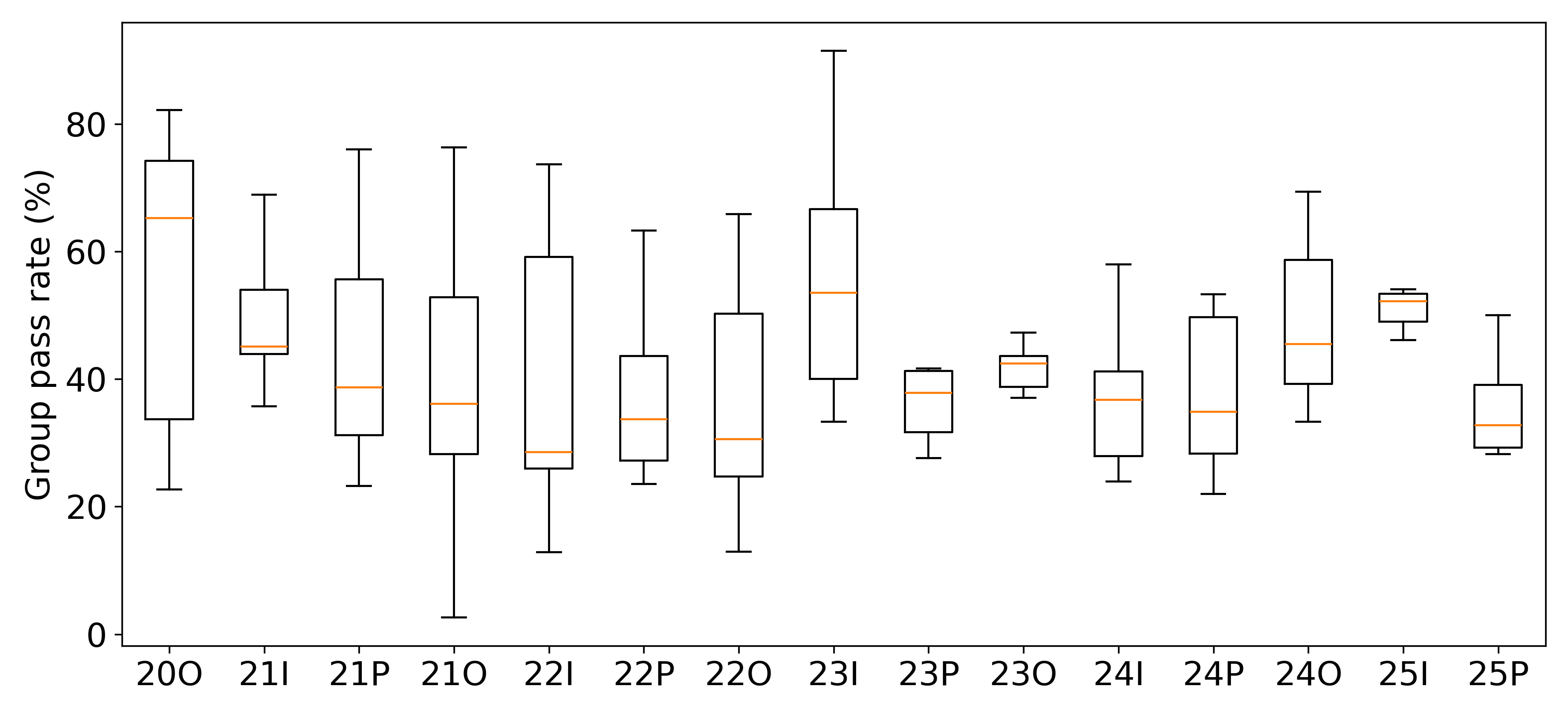}
\caption{Elementary Mechanics I: between-group variability in pass rates by term. Each box summarizes the distribution of group pass rates (marks greater than $6.0/10$) within a term.}
\label{fig:emi_tl_boxplot}
\end{figure}

To contextualize the magnitude and variability of student outcomes under the pre-intervention status quo, we summarize historical grade distributions for \emph{Elementary Mechanics I} (EM1) across terms 20-O (Autumn 2020) through 25-P (Spring 2025) (109 course groups; $N=4706$ enrolled students) in Table~\ref{tab:emi_tl_history} and Figure ~\ref{fig:emi_tl_boxplot}.  Because institutional conditions and policies (e.g., remote instruction periods) varied across 2020–2025, this historical distribution is used only as context rather than as a counterfactual for quasi-experimental comparison---though Figure~\ref{fig:emi_tl_boxplot} shows that the outcomes during the pandemic lockdowns (20-O to 21-O) were not necessarily outliers. To the best of our knowledge, these historical groups correspond to sections taught predominantly in a traditional-lecturing format; this characterization reflects departmental practice and collective instructor experience, since systematic classroom-observation records are not available for the historical period.

We define course passing as any mark above $6.0$ on a $10$-point scale; we report withdrawal rates separately and compute all percentages relative to enrolled students ($N$). This is roughly equivalent to examining ``DFW'' rates at a university in the United States. Across all terms, $6.18\%$ of enrolled students withdrew ($291/4706$), $49.53\%$ received did not pass ($2331/4706$), and $44.28\%$ passed the course ($2084/4706$). Conditioning on non-withdrawals yields a slightly higher pass rate of $47.20\%$ (2084 out of 4415 non-withdrawing students). Table~\ref{tab:emi_tl_history} reports these outcomes by term. Importantly, the historical baseline exhibits substantial between-group heterogeneity: the term-level pass rate ranges from $2.6\%$ to $91.5\%$ (in an individual group), with a median of $41.5\%$ and an interquartile range of $23.1$ percentage points (see Figure~\ref{fig:emi_tl_boxplot}). Withdrawal rates are typically small (median $3.3\%$) but can be extreme in some groups (up to $55\%$). These descriptive facts motivate treating baseline performance as a distribution rather than a single number, and they highlight the importance of accounting for section-to-section variability when comparing instructional approaches.

\begin{table}[t]
\centering
\small
\setlength{\tabcolsep}{5pt}
\begin{tabular}{lrrcccc}
\hline
Term & \#groups & $N$ & Withdraw (\%) & Not Pass (\%) & Pass (\%) & Pass (\%) excl.\ withdraw \\
\hline
20O & 7 & 321 & 0.0 & 44.2 & 55.8 & 55.8 \\
21I & 9 & 397 & 0.0 & 47.4 & 52.6 & 52.6 \\
21P & 6 & 278 & 0.0 & 54.7 & 45.3 & 45.3 \\
21O & 7 & 244 & 0.0 & 59.8 & 40.2 & 40.2 \\
22I & 9 & 364 & 0.0 & 57.7 & 42.3 & 42.3 \\
22P & 6 & 283 & 0.0 & 59.0 & 41.0 & 41.0 \\
22O & 7 & 352 & 0.0 & 54.0 & 46.0 & 46.0 \\
23I & 9 & 252 & 18.7 & 34.1 & 47.2 & 58.1 \\
23P & 6 & 276 & 7.2 & 42.4 & 50.4 & 54.3 \\
23O & 7 & 324 & 8.6 & 55.2 & 36.1 & 39.4 \\
24I & 9 & 376 & 7.7 & 54.8 & 37.5 & 40.6 \\
24P & 6 & 261 & 18.8 & 37.2 & 44.1 & 54.2 \\
24O & 7 & 330 & 6.4 & 43.9 & 49.7 & 53.1 \\
25I & 8 & 375 & 7.5 & 43.5 & 49.1 & 53.0 \\
25P & 6 & 249 & 10.0 & 58.2 & 31.7 & 35.3 \\
\hline
Total & 109 & 4706 & 6.2 & 49.5 & 44.3 & 47.2 \\
\hline
\end{tabular}
\caption{Historical outcomes for Elementary Mechanics I, aggregated by term. Percentages are computed relative to enrolled students ($N$). The ``Pass'' category aggregates marks greater than $6.0/10$.  The pass marks are labeled at our Institution as: \texttt{S} (sufficient), \texttt{B} (good), and \texttt{MB} (very good). ``Pass excl.\ withdraw'' conditions on students who did not withdraw.}
\label{tab:emi_tl_history}
\end{table}

In 2025, the department began a coordinated instructional change initiative to address the highly variable, but overall low, student pass rates. Though this change effort was started by I.P.C., it was not a ``top-down'' change effort  nor was it prescriptive \cite{henderson2011facilitating}. The effort instead made room for \emph{emergent} change in which instructors and teaching assistants worked together to develop a shared vision for effective teaching while still allowing individuals the agency to become reflective and innovative instructors. The approach most closely aligns with Borrego and Henderson's ``Complexity Leadership'' strategy \cite{borrego2014increasing}, in that the leader of the effort encouraged disruption of the status quo while simultaneously creating an environment in which faculty had agency to create their own changes as an interdependent collective. Though this paper is not designed to study the change initiative holistically, it represents an important artifact to amplify the initiative and the effective change that is happening as a result.

Two faculty members (L.J.L. and O.G.) in the Department of Physics volunteered (or agreed) to receive training in active-learning methods and to implement these methods in their sections for \emph{four} consecutive trimesters,  beginning in the trimester 25-O. From the outset they were informed that initial implementation would require a steep learning curve, substantial preparation of instructional materials, and a change in instructional mindset relative to traditional lecturing. Four teaching assistants (TAs) were selected to participate in the same effort and received the same expectations regarding preparation, coordination, and sustained implementation. This approach drew on literature highlighting the need for \emph{sustained} efforts \cite{henderson2011facilitating,wieman2017improving} to sustain cultural change in instruction.

Preparation for the active-learning implementation occurred through a sequence of workshops and structured follow-up work. The initial training was a five-day workshop led by E.B. in June 2025. This workshop was not a targeted training in a specific active-learning technique, but rather a hands-on application of canonical theories of learning rooted in decades of empirical findings---inspired largely by the Science Education Initiative \cite{wieman2017improving}. Though fidelity of implementation has been a concern in the literature on instructional change \cite{WiemanPerkins2005}, there are more recent and detailed studies which suggest that different ways of implementing critical components of active learning are just as effective as implementing specific named methods like Peer Instruction \cite{bukola2025measuring}.

This was followed by three meetings in July focused on course content and on articulating and classifying learning objectives (including explicit discussion of their placement within Bloom's Taxonomy \cite{bloom1956taxonomy}), with the goal of producing clear descriptive letters and learning-goals statements to guide alignment between in-class activities and targeted outcomes. A WhatsApp group was created for the two trained instructors and four participating TAs, and during the summer break (August/September) this group coordinated the preparation of instructional materials and active-learning activities informed by the June workshop and guided by the learning-goals documents. A second workshop (two days) was delivered by a different renowned expert in active learning to consolidate the team’s training, discuss implementation choices, and review the suitability of the materials prepared during the summer period right before term 25-O began. In parallel, two educational psychologists were hired and trained (with guidance from E.B.) to use the Classroom Observation Protocol for Undergraduate STEM (COPUS) as an observational tool \cite{Smith2013COPUS}. This was chosen specifically because we were not encouraging prescribed instructional changes, so we wanted to have more concrete data on how these instructors were implementing AL.

During the trimester 25-O, the two trained instructors each taught one section of Elementary Mechanics I. Each of these sections was assigned two TAs, reflecting the additional logistical demands of implementing active-learning activities and preparing aligned materials for the first time. The remaining six sections were taught by six instructors who had taught the course multiple times; those sections were each assigned one TA, maintaining established instructor--TA pairings when feasible. It is relevant for interpreting results that the two instructors leading the active-learning sections had not taught Elementary Mechanics I for some time, so in terms of course-specific familiarity they began at a disadvantage relative to instructors who taught the course repeatedly.

The trimester 25-O ran from 1 October 2025 to 12 December 2025. During ``week zero'', all instructors and TAs across the eight sections received information on the official course content, and the descriptive letters were shared with all sections to communicate the intent of defining clear learning goals and aligning activities with those goals. During the trimester, the coordinator of the Physics core curriculum (I.P.C.) monitored the progress of all sections, particularly around examinations, and intervened when anomalous performance was detected by requesting explanations and asking for corresponding adjustments in instructional activities. In practice, such interventions occurred in some TL-labeled sections when unexpectedly poor performance was observed; after the intervention, performance improved on the second midterm examination.

To document classroom practices and provide qualitative evidence of implementation, the two educational psychologists conducted COPUS observations during weeks 2--3 and again during weeks 8--9. Observations were carried out in the two active-learning sections and in two traditional-lecturing sections; instructors were informed that classroom monitoring would occur during the term, but neither the specific observation dates nor the observation instrument were disclosed in advance. In addition to these structured observations, the coordinator maintained regular communication with the active-learning instructors and TAs through group meetings, individual meetings, and the WhatsApp group used for day-to-day coordination. A third workshop, led by E.B., was held in November 2025 as a continuation of the earlier training. During this visit, E.B. attended class sessions taught by the trained instructors and the TAs and provided direct, personalized feedback to both instructors and TAs on their active-learning implementation using the Reformed Teaching Observation Protocol \cite{rtop}. Throughout the term, substantial instructional materials were developed and refined, including question banks, clicker/slicker questions, and worksheets; a large fraction of this material was prepared during the summer break, with additional material developed during the trimester in response to pacing and implementation needs.

For the purposes of this study, sections are classified as Active Learning (AL) if they were taught by the two instructors and supported by the TAs who received the active-learning training described above, and if they implemented active-learning methods as monitored during the term. The remaining sections are classified as Traditional Lecturing (TL) because their instructors did not participate in the training sequence. We emphasize that this classification is operational and reflects the initiative’s implementation: some instructors in the TL-labeled sections may have occasionally used activities resembling active-learning techniques, either informally or through prior exposure, but without participating in the structured preparation and support described above.

Common examinations were coordinated at the course level. Approximately one week before each midterm examination and the global examination, the coordinator convened all instructors (both AL and TL) to prepare a common exam for all sections based on the material covered. In contrast, the remaining $50\%$ of the final mark was allocated by each instructor according to their own assessment design. Instructors participating in the active-learning initiative typically used assessment components aligned with active-learning practice (e.g., participation-oriented or formative elements), whereas instructors in TL-labeled sections made independent choices within the same institutional policy. Finally, implementation fidelity during the term was supported qualitatively through frequent discussions with the trained instructors and TAs to ensure that active-learning-aligned activities were being used and that prepared materials were being deployed as intended.

%%%%%%%%%%%%%%%%%%%%%%%%%%%%%%%%%%%%%%%%%%%%%%%%%%%%%%%%%%%%%%%%%%%%%%%%%%%%%%
\section{Data Collected and Caveats}
\label{sec:data}
%%%%%%%%%%%%%%%%%%%%%%%%%%%%%%%%%%%%%%%%%%%%%%%%%%%%%%%%%%%%%%%%%%%%%%%%%%%%%%
The study was conducted in \textit{Elementary Mechanics I} during term 25-O. The trimester follows the institutional format of 11 weeks of instruction with 6 contact hours per week. The course includes two midterm examinations (administered in weeks 5 and 9) and a common global examination (administered after last week of classes, locally so-called week 12). These three exams jointly account for 50\% of the final recorded course grade (final mark). The remaining 50\% of the final mark is determined at the instructor's discretion through additional activities and evaluations. In term 25-O the course was offered in eight sections scheduled throughout the day, each with a nominal capacity of approximately 50 students. Students enroll by selecting a section based on their preferences and scheduling constraints. No prior announcement was given to students about the active learning initiative.

The dataset used in this paper consists exclusively of aggregated counts extracted from standardized end-of-term group reports for \textit{Elementary Mechanics I} in term 25-O, prepared by the course coordinator and completed by the TAs. For each section and each common departmental assessment (first and second midterm exams and the global exam), the official report provides--among other things--the number of students who took the assessment and the number who passed. For the final mark, the report is organized around recorded final grades (not exam attendance). We therefore treat as the relevant count the number of students with a recorded final grade, and we count as passing those students whose recorded final grade corresponds to passing the course. Student-level grades, item-level exam scores, demographic variables, prior preparation indicators, and imputed values are not used in the analysis. All computations reported in this paper are functions of these section-level counts only, consistent with the scope and constraints described in the Introduction.

Table~\ref{tab:data} lists the complete set of per-section inputs used in the analysis, including the AL/TL classification. These inputs are aggregated into pooled totals by modality and assessment to estimate pooled student-level pass probabilities that we will define precisely in the methods section. When a section has a missing entry for a particular assessment in the official report (e.g., the global exam is not recorded for one section), that section is excluded from the pooled totals for that assessment only, while remaining included for assessments where the report provides the relevant counts.

\begin{table}[h!]
 \begin{tabular}{l l r r r r r r r r}
\toprule
Section & Mode & \multicolumn{2}{c}{1st midterm exam} & \multicolumn{2}{c}{2nd midterm exam} & \multicolumn{2}{c}{Global exam} & \multicolumn{2}{c}{Final mark} \\
\cmidrule(lr){3-4}\cmidrule(lr){5-6}\cmidrule(lr){7-8}\cmidrule(lr){9-10}
   &  & Took & Pass & Took & Pass & Took & Pass & Recorded & Pass \\
\midrule
 \#1 & TL & 37 &  9 & 35 & 18 & 34 &  8 & 46 & 20 \\
 \#2 & AL & 42 & 19 & 39 & 19 & 36 & 19 & 45 & 23 \\
 \#3 & AL & 47 & 24 & 45 & 16 & 45 & 26 & 51 & 42 \\
 \#4 & TL & 50 &  1 & 50 &  9 & 41 &  6 & 57 & 15 \\
 \#5 & TL & 32 & 20 & 32 & 12 & 30 & 15 & 39 & 13 \\
 \#6 & TL & 45 & 21 & 46 & 16 & 48 & 21 & 50 & 39 \\
 \#7 & TL & 21 & 10 & 17 & 10 & -- & -- & 24 & 16 \\
 \#8 & TL & 40 & 20 & 37 & 12 & 37 & 12 & 37 & 11 \\
\bottomrule
\end{tabular}   

\caption{Per-group counts used in the analysis. ``Took'' is the number of students who took the assessment; ``Pass'' is the number who passed. For the final mark, ``Recorded'' is the number of students with a recorded final grade and ``Pass'' is the number who passed the course.}
\label{tab:data}
\end{table}

In addition to outcome data, the initiative included limited observational evidence regarding classroom practices. Because the labels AL and TL in this study are operational---they refer to how sections were implemented in practice rather than to a single standardized instructional ``treatment''---we include COPUS observations as structured, qualitative support for the intended meaning of these labels. As described in Section~\ref{sec:initiative}, two trained observers (M.A.M.O. and E.B.U.) used the Classroom Observation Protocol for Undergraduate STEM (COPUS) to document instructor and student behaviors in a small sample of class meetings. Observations were conducted during weeks 2--3 and 8--9 in the two AL-labeled sections and in two TL-labeled sections. Instructors were aware that classroom monitoring would occur during the term, but they were not informed of the specific observation dates nor of the observation instrument. COPUS is not used to compute a quantitative fidelity score and does not enter the statistical model for pass-rate outcomes; instead, it provides implementation evidence that the AL and TL labels correspond to meaningfully different patterns of classroom activity in this setting.

\begin{figure*}[t]
\centering
\includegraphics[width=\textwidth]{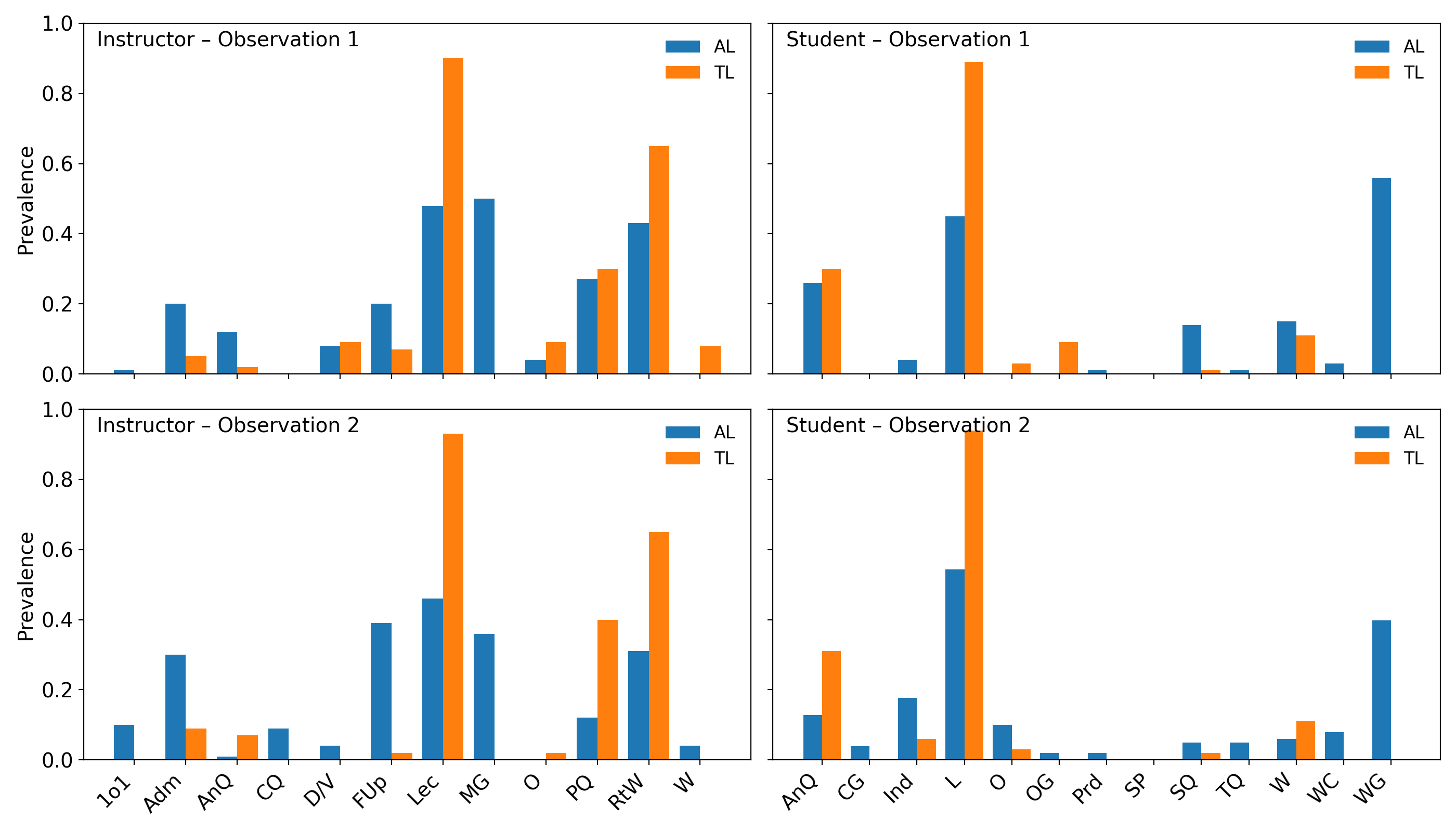}
\caption{COPUS prevalence (fraction of 2-min intervals) by modality (AL vs TL) for two observation rounds, shown separately for student and instructor codes.}
\label{fig:copus-prevalence}
\end{figure*}

\begin{table*}[t]
\centering
\begin{threeparttable}
\begin{tabular}{lcccc}
\toprule
COPUS code  & Obs.~1 AL & Obs.~1 TL & Obs.~2 AL & Obs.~2 TL \\
\midrule
\multicolumn{5}{l}{\textbf{Students}}\\
\midrule
AnQ (Answering a question)   & 0.260 & 0.300 & 0.129 & 0.310 \\
CG (Clicker-group discussion)   & 0.000 & 0.000 & 0.039 & 0.000 \\
Ind (Individual thinking/working)   & 0.040 & 0.000 & 0.177 & 0.060 \\
L (Listening)   & 0.450 & 0.890 & 0.543 & 0.940 \\
O (Other)   & 0.000 & 0.030 & 0.100 & 0.030 \\
OG (Other group activity)   & 0.000 & 0.090 & 0.020 & 0.000 \\
Prd (Prediction)   & 0.010 & 0.000 & 0.020 & 0.000 \\
SP (Student presentation)   & 0.000 & 0.000 & 0.000 & 0.000 \\
SQ (Student question)   & 0.140 & 0.010 & 0.049 & 0.020 \\
TQ (Test/quiz)   & 0.010 & 0.000 & 0.049 & 0.000 \\
W (Waiting)   & 0.150 & 0.110 & 0.059 & 0.110 \\
WC (Whole-class discussion)   & 0.030 & 0.000 & 0.078 & 0.000 \\
WG (Working in groups)   & 0.560 & 0.000 & 0.397 & 0.000 \\
\midrule
\multicolumn{5}{l}{\textbf{Instructors}}\\
\midrule
1o1 (One-on-one)   & 0.010 & 0.000 & 0.100 & 0.000 \\
Adm (Administration)  & 0.200 & 0.050 & 0.300 & 0.090 \\
AnQ (Answering a question)   & 0.120 & 0.020 & 0.010 & 0.070 \\
CQ (Clicker question)   & 0.000 & 0.000 & 0.090 & 0.000 \\
D/V (Demonstration / video)   & 0.080 & 0.090 & 0.040 & 0.000 \\
FUp (Follow-up / feedback)   & 0.200 & 0.070 & 0.390 & 0.020 \\
Lec (Lecturing)   & 0.480 & 0.900 & 0.460 & 0.930 \\
MG (Moving/guiding; monitoring groups)   & 0.500 & 0.000 & 0.360 & 0.000 \\
O (Other)   & 0.040 & 0.090 & 0.000 & 0.020 \\
PQ (Posing a question)   & 0.270 & 0.300 & 0.120 & 0.400 \\
RtW (Real-time writing)  & 0.430 & 0.650 & 0.310 & 0.650 \\
W (Waiting) & 0.000 & 0.080 & 0.040 & 0.000 \\
\bottomrule
\end{tabular}
\begin{tablenotes}[flushleft]
\footnotesize
\item \textit{Notes.} Prevalence is defined as the number of 2-min intervals in which a code occurs divided by the total number of 2-min intervals observed. Because multiple codes can be recorded in the same interval, prevalences across codes do not sum to 1. ``Other'' categories (O) should be interpreted cautiously because their content depends on local coding notes.
\end{tablenotes}
\caption{COPUS prevalence (fractions of 2-min intervals) for all coded behaviors, by observation round and modality. Values are aggregated within modality across observed sections in each round.}
\label{tab:copus-allcodes}
\end{threeparttable}
\end{table*}

\begin{table}[t]
\centering
\small
\setlength{\tabcolsep}{4pt}
\begin{tabular}{llrr}
\toprule
Role & Code & Obs.~1 $\Delta$ & Obs.~2 $\Delta$ \\
\midrule
Students & WG  & +0.560 & +0.397 \\
Students & L   & -0.440 & -0.397 \\
Students & AnQ & -0.040 & -0.181 \\
\midrule
Instructors & MG   & +0.500 & +0.360 \\
Instructors & Lec  & -0.420 & -0.470 \\
Instructors & RtW  & -0.220 & -0.340 \\
Instructors & FUp  & +0.130 & +0.370 \\
Instructors & PQ   & -0.030 & -0.280 \\
\bottomrule
\end{tabular}
\caption{Key COPUS prevalence contrasts (AL$-$TL) for selected codes. Prevalence is the fraction of 2-min intervals in which a code occurs; positive values indicate higher prevalence in AL.}
\label{tab:copus-key}
\end{table}

We report COPUS results using \emph{prevalence}, defined as the fraction of 2-min intervals in which a code is observed (Fig.~\ref{fig:copus-prevalence}). Because multiple behaviors can co-occur within the same interval, prevalence across codes does not necessarily sum to 1. Figure~\ref{fig:copus-prevalence} summarizes prevalence by modality for students  and instructors; Table~\ref{tab:copus-allcodes} reports the complete prevalence values for all coded behaviors.

The clearest modality contrasts are summarized in Table~\ref{tab:copus-key}. Across both observation rounds, the most salient student-side contrast is the presence of group work in AL (WG) and its absence in the observed TL meetings. On the instructor side, AL shows higher prevalence of facilitation behaviors (MG and FUp), whereas TL remains more lecture- and board-centered (Lec and RtW). Notably, the largest and most reproducible contrasts across rounds are the near-absence of student group work in TL and the corresponding shift from instructor lecturing/writing (Lec, RtW) to instructor facilitation (MG, FUp) in AL. Taken together, these COPUS profiles provide structured support that the AL/TL labels used in the outcome analysis corresponded to meaningfully different classroom interaction patterns in the observed meetings, while remaining limited in scope due to the small number of observed sections and class meetings sampled.

Several caveats are essential for interpreting the results. First, the core curriculum coordinator’s role cannot be passive. When anomalous performance is detected in any section, the coordinator must intervene as part of standard academic oversight, independent of the active-learning initiative. In term 25-O, such interventions occurred in some TL-labeled sections after unexpectedly low pass rates were observed, and the instructors and TAs were asked to explain the situation and adjust instructional activities accordingly. This means that the study should not be interpreted as a controlled comparison of two fixed instructional regimes; it reflects ``AL as implemented'' and ``TL as implemented'' under normal departmental governance, including corrective actions when warranted.

Second, instructors in TL-labeled sections generally had recent and accumulated experience teaching \textit{Elementary Mechanics I}, whereas the instructors leading the AL implementation had not taught this specific course for some time. This difference in course-specific familiarity is part of the real implementation context and may work against observing an AL advantage during early adoption. More broadly, the assignment of instructors to sections was not randomized and students self-selected into sections based on scheduling and personal preferences. Without student-level covariates, the analysis cannot adjust for potential differences in student composition across sections, and the study does not claim causal identification of an instructional ``treatment effect''. Instead, the analysis provides a transparent, student-weighted benchmark of pooled passing outcomes under a shared assessment standard for the two operationally defined modalities, precisely because it reflects a real-world departmental implementation carried out with the limited and realistic instructional supports typically available when initiatives of this kind are adopted.

Finally, the use of aggregated counts implies that standard statistical models (i.e., logistic regression or $\chi^2$ tests) treat student outcomes as independent Bernoulli trials within each modality and assessment, even though outcomes may be correlated within sections. Such clustering, if present, could reduce effective sample sizes relative to naive binomial assumptions \cite{van2019modernizing}. This limitation motivates the conservative interpretation adopted in this paper: the estimates quantify differences in pooled pass rates and their associated uncertainty under a working model, while the discussion and conclusions emphasize what can and cannot be inferred from aggregated institutional data.

This study uses aggregated, de-identified section-level counts from routine course reporting; no student-level records were accessed. Under The Declaration of Helsinki for research ethics review, this analysis was determined to be exempt from human-subjects review because it involves only non-identifiable, aggregate administrative and observational data.

%%%%%%%%%%%%%%%%%%%%%%%%%%%%%%%%%%%%%%%%
\section{Statistical Analysis: Methods and Results}
\label{sec:analysis}
%%%%%%%%%%%%%%%%%%%%%%%%%%%%%%%%%%%%%%%%
%%%%%%%%%%%%%%%%%%%%%%%%%%%%%%%%%%%%%%%
\subsection{Methods}
%%%%%%%%%%%%%%%%%%%%%%%%%%%%%%%%%%%%%%%
We compare student passing outcomes between sections classified as Active Learning (AL) and Traditional Lecturing (TL) using a pooled, student-level analysis based on the common departmental assessments. It is important to emphasize what this comparison represents: the labels AL and TL refer to the way each section was implemented in practice, and individual instructors may employ different mixes of instructional techniques within the same label. Accordingly, the pooled estimates should be interpreted as the average pass probabilities
\begin{equation}
    \begin{split}
 p_{\beta,a}=\text{Prob}(\text{pass mark for assessment }a \mid &\text{student has an observed outcome for }a\\
 &\text{ in a }\beta\text{-labeled section})\,,
    \end{split}
\end{equation}
where the conditioning is on the section label as \(\beta\in\{\mathrm{AL},\mathrm{TL}\}\) as implemented and on the specific assessment \(a\in\{\)first midterm exam, second midterm exam, global exam, final mark\(\}\) common to all sections, with the caveat explained before for the final mark. This provides a transparent benchmark of \textit{AL as implemented} versus \textit{TL as implemented} under a shared assessment standard, with sections contributing in proportion to their student counts; the final mark is reported as an additional outcome but is only partially standardized. However, it should not be interpreted as isolating the causal effect of any single instructional technique, nor as the effect of a perfectly standardized teaching protocol. This choice reflects a pragmatic evaluation goal aligned with departmental decision-making. Indeed, even though pass rates are an incomplete proxy for learning and can be influenced by factors beyond conceptual mastery, they remain a consequential outcome in gateway courses and, when interpreted carefully under common assessments, can provide an informative first benchmark that is feasible to compute and reproduce.

The practical question behind this analysis is straightforward: for each assessment in the course, do students in AL-labeled sections pass at a higher rate than students in TL-labeled sections, and if so, by how much? The outputs of interest are therefore (i) pass-rate estimates for each modality, (ii) an effect size on an interpretable scale, and (iii) uncertainty quantification.

For each assessment \(a\in\{\)first midterm exam, second midterm exam, global exam, final mark\(\}\), we focus on the two underlying pass probabilities $p_{\mathrm{AL},a}$ and $p_{\mathrm{TL},a}$. The primary estimand for comparing teaching modalities is the  risk difference in passing between both modalities, that is:
\begin{equation}
\mathrm{RD}_a \;=\; p_{\mathrm{AL},a} - p_{\mathrm{TL},a}\,,
\label{eq:RD}
\end{equation}
reported in percentage points. This is the most directly interpretable scale for instructional decisions: it answers how many additional students pass per 100 under AL with respect to TL for each assessment.

The dataset consists exclusively of aggregated counts (see Table~\ref{tab:data} in Section~\ref{sec:data}). For a fixed assessment $a$ and modality $\beta$, let $X_{\beta,a}$ be the total number of $\beta$-labeled students  who passed and let $N_{\beta,a}$ be the total number of $\beta$-labeled  students observed for that assessment; both totals are obtained by summing the corresponding group counts within $\beta$-labeled  students.  Pooling is the primary analysis because the comparison is intended to be student-level, not section-level, since a section-level average would treat a section of 30 students and a section of 60 students as equally informative, even though the latter contains roughly twice as many Bernoulli outcomes.

Now, within each modality and assessment, each student outcome is treated as a Bernoulli random variable with pass probability $p_{\beta,a}$. Thus pooling yields the binomial model
\begin{equation}
X_{\beta,a}\sim \mathrm{Binomial}(N_{\beta,a}, p_{\beta,a}),\qquad \beta\in\{\mathrm{AL},\mathrm{TL}\}\,,
\label{eq:pooledmodel}
\end{equation}
where  student assessment outcomes are treated as independent and identically distributed Bernoulli random variables. It captures the dominant sampling uncertainty associated with the finite sample size $N_{\beta,a}$, while acknowledging that real courses may exhibit additional structure (e.g., within-section correlation and section-level heterogeneity). We address this limitation explicitly below by conducting a sensitivity analysis using a hierarchical model designed to accommodate clustering. The implications of aggregation and potential within-section dependence are also discussed in the interpretation boundary for the results.

Under the binomial model, the pooled pass-rate frequentist estimators of the probability $p_{\beta,a}$ are therefore given by:
\begin{equation}
\widehat{p}_{\beta,a}=\frac{X_{\beta,a}}{N_{\beta,a}},\qquad \beta\in\{\mathrm{AL}, \mathrm{TL}\}\,.
\end{equation}
To quantify the uncertainty of the estimators $\widehat{p}_{\beta,a}=\frac{X_{\beta,a}}{N_{\beta,a}}$, we use Wilson's binomial confidence intervals with score $z=1.96$ to report 95\%  confidence interval, so that  Wilson's interval endpoints $p^{(\pm)}_{\beta,a}$ are given by
\begin{equation}
\label{eq:wilson}
p^{(\pm)}_{\beta,a}=\frac{\widehat p_{\beta,a} + \frac{z^2}{2N_{\beta,a}}\pm z\sqrt{\frac{\widehat p_{\beta,a}(1-\widehat p_{\beta,a})}{N_{\beta,a}}+\frac{z^2}{4N^2_{\beta,a}}}}{1+\frac{z^2}{N_{\beta,a}}}\,.
\end{equation}
We decided to use Wilson's intervals \cite{wilson1927probable} as they typically achieve more reliable coverage for moderate sample sizes and for proportions near 0 or 1, which makes them well suited for pass-rate reporting.

To better appreciate the effect size between modalities we use the frequentist estimator to the risk difference as
\begin{equation}
\widehat{\mathrm{RD}}_a=\widehat p_{\mathrm{AL},a}-\widehat p_{\mathrm{TL},a}\,.
\label{eq:RDestimator}
\end{equation}
Furthermore, to  quantify uncertainty for this risk difference estimator we avoid the naive Wald interval for a difference of proportions and instead use a score-based interval consistent with the Wilson intervals used for each pooled proportion. Let $p^{(\pm)}_{\beta,a}$ denote the Wilson 95\% endpoints for $p_{\beta,a}$ from Eq.~\eqref{eq:wilson}. We report the 95\% confidence interval for $\mathrm{RD}_a=p_{\mathrm{AL},a}-p_{\mathrm{TL},a}$ as the Newcombe--Wilson interval
\begin{equation}
\mathrm{CI}_{95\%}(\mathrm{RD}_a)=\Big[p^{(-)}_{\mathrm{AL},a}-p^{(+)}_{\mathrm{TL},a}\,,\; p^{(+)}_{\mathrm{AL},a}-p^{(-)}_{\mathrm{TL},a}\Big],
\label{eq:RD_newcombe}
\end{equation}
which has more reliable finite-sample behavior than the Wald normal approximation, particularly when pass rates are far from $0.5$.

To complement this frequentist presentation---and to set up the hierarchical sensitivity analysis below---we also adopt a Bayesian formulation of the same pooled binomial model. This is particularly convenient here because we are estimating pass probabilities constrained to $[0,1]$, sample sizes are moderate, and some assessments exhibit pass rates sufficiently far from 0.5 that naive normal approximations can be unreliable. In a Bayesian analysis, uncertainty in each $p_{\beta,a}$ is represented directly by a posterior distribution, yielding finite-sample credible intervals and decision-relevant summaries such as $\text{Prob}(\mathrm{RD}_a>0\mid\text{data})$. This pooled Bayesian specification is intentionally minimal and can be extended in future terms (e.g., informative priors or richer hierarchical structure) as additional evidence accumulates. Moreover, the Bayesian framework provides a natural bridge to the hierarchical sensitivity analysis discussed above, in which section-level heterogeneity and within-section dependence can be modeled explicitly via partial pooling. In this example, we can use a section-level estimation of passing probabilities to inform the estimations of the pooled student-level passing probabilities to account for potential section-level impacts on average performance which may be conflated with the frequentist estimates.

More precisely, we first assume a weakly noninformative Jeffreys prior on the pass probability given by $p_{\beta,a}\sim \mathrm{Beta}(\tfrac12,\tfrac12)$. When combining  the binomial likelihood with the prior, this yields  a posterior distribution of the pass rates that also belongs, by prior conjugacy, to the Beta distribution family,
\begin{equation}
  p_{\beta,a}\mid X_{\beta,a},N_{\beta,a}\sim
\mathrm{Beta}\!\left(X_{\beta,a}+\tfrac12,\;N_{\beta,a}-X_{\beta,a}+\tfrac12\right)\,.
\end{equation}
From these posteriors we report posterior means and central 95\% credible intervals for each \(p_{\beta,a}\). The posterior distribution of \(\mathrm{RD}_a\) is, in turn, estimated by Monte Carlo sampling, which yields posterior means, 95\% credible intervals, and the posterior probability \(\text{Prob}(\mathrm{RD}_a>0\mid\text{data})\). 

The pooled pass binomial model in Eq.~\eqref{eq:pooledmodel} treats all student outcomes within a given modality and assessment as independent Bernoulli trials with a common pass probability $p_{\beta,a}$. This yields a transparent reference analysis from the aggregated totals $(X_{\beta,a},N_{\beta,a})$, but it may understate uncertainty when outcomes are correlated within course sections (i.e., when there is clustering). To assess the robustness of our conclusions to such clustering using only the available section-level counts (Table~\ref{tab:data}), we fit a section-clustered hierarchical model and, crucially, map its section-level predictions back to the same pooled, student-weighted estimands $p_{\mathrm{AL},a}$ and $p_{\mathrm{TL},a}$ defined in Eq.~\eqref{eq:pooledmodel}, together with the corresponding risk difference $\mathrm{RD}_a$ in Eq.~\eqref{eq:RD}. More precisely, let $\ell$ index sections, and let $X_{\ell,a}$ and $N_{\ell,a}$ denote, respectively, the number of students who pass and the number observed in section $\ell$ for assessment $a$ (for the final mark, $N_{\ell,a}$ is the number of recorded final grades, as in Table~\ref{tab:data}). We model within-section outcomes as
\begin{equation}
X_{\ell,a}\sim \mathrm{Binomial}(N_{\ell,a},p_{\ell,a}),
\label{eq:hier_binom}
\end{equation}
where $p_{\ell,a}$ is the section-specific pass probability for assessment $a$. To represent persistent differences between sections (e.g., instructor/implementation and section-level composition effects) we use a logistic mixed-effects specification,
\begin{equation}
\mathrm{logit}(p_{\ell,a})=\alpha_a + u_\ell + \gamma_a\,\mathbb{I}\{\ell\in \mathrm{AL}\},
\label{eq:hier_logit}
\end{equation}
where $\alpha_a$ is an assessment baseline (overall difficulty/stringency), $u_\ell$ is a section random intercept capturing between-section heterogeneity, and $\gamma_a$ is the assessment-specific AL--TL shift on the log-odds scale (so that $\exp(\gamma_a)$ is the corresponding odds ratio conditional on $u_\ell$). We complete the Bayesian specification using the default weakly informative priors of the fitting procedure. Specifically, we place independent normal priors on all fixed-effect coefficients (both $\alpha_a$ and $\gamma_a$),  $\alpha_a,\gamma_a \sim \mathcal{N}(0,\,2^2)$
and model section effects as $u_\ell\mid\sigma \sim \mathcal{N}(0,\sigma^2)$ with a normal prior on the log standard deviation, $\log \sigma \sim \mathcal{N}(0,\,1^2)$.
These priors, and the values chosen for the hyperparameters, are proper and weakly regularizing on the log-odds scale, discouraging extreme probabilities unless strongly supported by the binomial counts.

Crucially, the pooled pass probabilities reported previously remain student-weighted quantities. Given a posterior draw of the section-level probabilities $\{p_{\ell,a}\}$, we define the implied pooled pass probability for modality $\beta\in\{\mathrm{AL},\mathrm{TL}\}$ as the student-weighted average
\begin{equation}
p_{\beta,a}=\sum_{\ell\in\beta} w_{\ell,a}\,p_{\ell,a}\,,\qquad
w_{\ell,a}=\frac{N_{\ell,a}}{\sum_{\ell'\in\beta} N_{\ell',a}}=\frac{N_{\ell,a}}{N_{\beta,a}}\,.
\label{eq:hier_pool}
\end{equation}
This mapping ensures that $p_{\beta,a}$ targets the same pooled student-level estimand as in the binomial model given by Eq.~\eqref{eq:pooledmodel}: sections contribute in proportion to their student counts. If a section has a missing entry for assessment $a$ in the official report, it is omitted from the corresponding sum in Eq.~\eqref{eq:hier_pool} for that assessment only, consistent with the pooling rule stated previously.

Posterior uncertainty for $p_{\beta,a}$ and for $\mathrm{RD}_a=p_{\mathrm{AL},a}-p_{\mathrm{TL},a}$ is obtained by first approximating the joint posterior distribution of the model parameters $(\alpha_a,\gamma_a,u_\ell,\sigma)$ with a variational Bayes procedure, and then drawing Monte Carlo samples from this approximate posterior. From these draws we summarize $p_{\beta,a}$ and $\mathrm{RD}_a$ by reporting posterior means and central 95\% credible intervals, together with the posterior probability $\text{Prob}(\mathrm{RD}_a>0\mid \mathrm{data})$.

In the Results subsection below we report (i) pooled totals by modality and assessment, (ii) pooled pass-rate estimates with Wilson intervals, (iii) risk differences with Newcombe–Wilson intervals, (iv) Bayesian posterior summaries under the pooled model, and (v) a hierarchical sensitivity analysis that relaxes independence via a section random intercept.

%%%%%%%%%%%%%%%%%%%%%%%%%%%%%%%%
\subsection{Results}
%%%%%%%%%%%%%%%%%%%%%%%%%%%%%%%%
Because the analysis is purely count-based, it is useful to present the pooled totals $(X_{\beta,a},N_{\beta,a})$ underlying each pooled proportion. This is reported in  Table~\ref{tab:pooled-totals} whose values fully determine the pooled pass estimates and their binomial uncertainties.
\begin{table}[h!]
\centering
 \begin{tabular}{l r r r r}
\toprule
Assessment & \multicolumn{2}{c}{Active Learning (AL)} & \multicolumn{2}{c}{Traditional Lecturing (TL)} \\
\cmidrule(lr){2-3}\cmidrule(lr){4-5}
 & Pass $X$ & Took $N$ & Pass $X$ & Took $N$ \\
\midrule
First midterm exam & 43 &  89 &  81 & 225 \\
Second midterm exam & 35 &  84 &  77 & 217 \\
Global exam & 45 &  81 &  62 & 190 \\
Final mark & 65 &  96 & 114 & 253 \\
\bottomrule
\end{tabular}   
\caption{Pooled totals by modality and assessment. ``Pass'' is $X$ and ``Took'' (or ``Recorded'' as it must be understood for the final mark) is $N$.}
\label{tab:pooled-totals}
\end{table}

\begin{figure}[h!]
\centering
\includegraphics[width=0.95\textwidth]{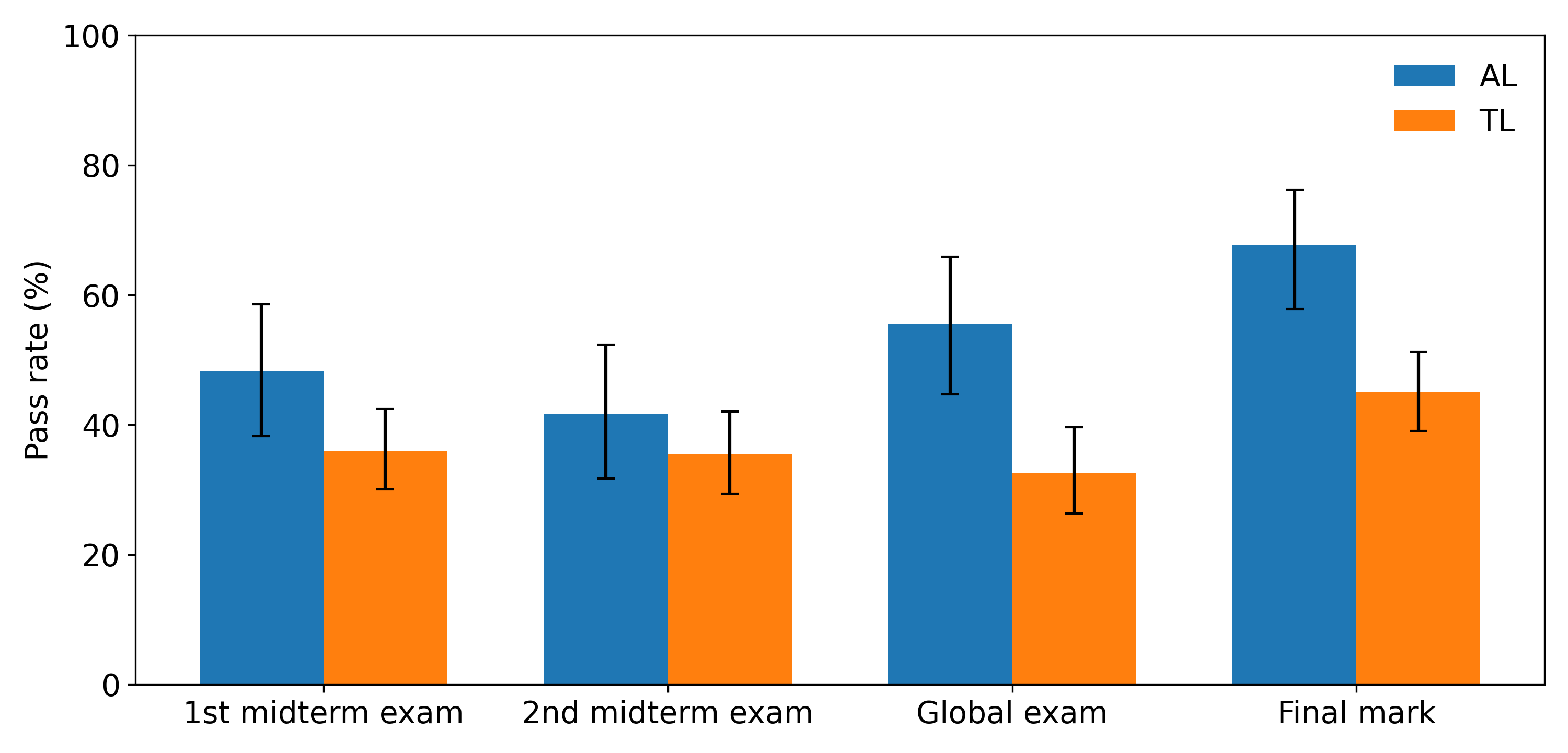}    
\caption{Pooled pass rates for Active Learning (AL) and Traditional Lecturing (TL) by assessment. Error bars represent 95\% binomial Wilson confidence intervals for each pooled proportion.}
\label{fig:pooled}
\end{figure}

Figure~\ref{fig:pooled} visualizes the pooled pass-rate estimators for AL and TL by assessment. Bar heights represent the pooled estimates $\widehat{p}_{\beta,a}$, and the error bars show 95\% Wilson confidence intervals for the corresponding pass probability in each modality. These intervals quantify uncertainty for each modality \emph{separately}. Accordingly, overlap (or non-overlap) of the two sets of error bars is not, by itself, a decision rule for the magnitude or statistical support of an AL--TL difference. To compare modalities directly, we therefore report uncertainty for an effect size computed from both pooled proportions. Descriptively, the pooled point estimates are higher for AL than for TL across the common assessments.

\begin{figure}[h!]
\centering
\includegraphics[width=0.95\textwidth]{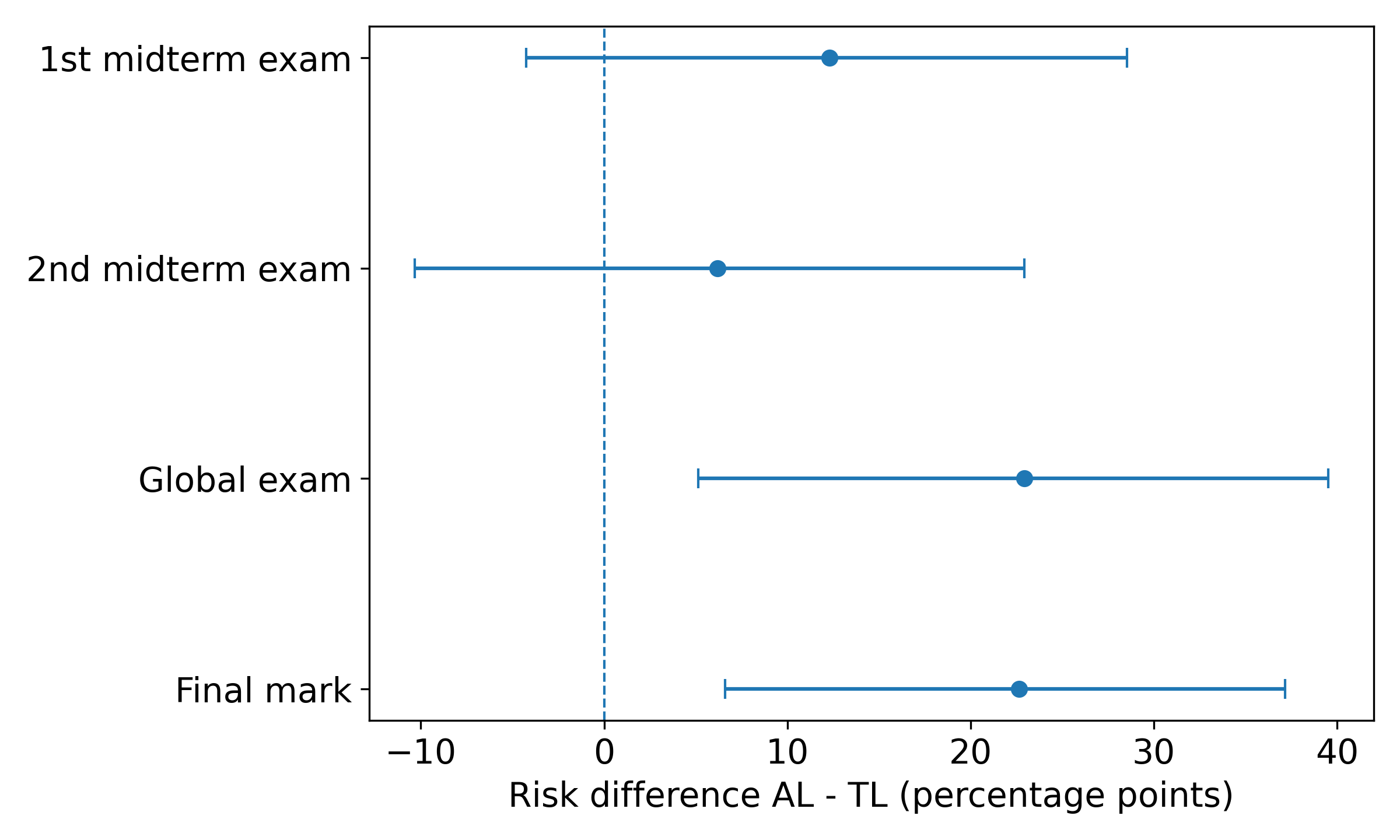}    
\caption{Risk difference between modalities (AL$-$TL) in percentage points with 95\% confidence intervals. The dashed line at 0 indicates no difference.}
\label{fig:forest}
\end{figure}

The uncertainty of the risk difference is presented in Figure~\ref{fig:forest}. Each point corresponds to the risk difference $\widehat{\mathrm{RD}}$ (AL minus TL) in percentage points, and the horizontal interval is the 95\% confidence interval for that difference. The dashed vertical reference at zero represents no difference. If the difference interval crosses zero, the data are compatible with both a positive effect and no effect at the 95\% level under the stated approximation. If the interval lies entirely above zero, the results support a positive AL advantage for that assessment at that confidence level. Table~\ref{tab:pooled-summary} reports the corresponding pooled pass rates and the risk difference.

\begin{table}[h!]
\centering
   \begin{tabular}{l
                S[table-format=2.1]
                S[table-format=2.1]
                S[table-format=2.1]
                l}
\toprule
Assessment & {AL pass} & {TL pass} & {RD (pp)} & {95\% CI for RD (pp)} \\
\midrule
First midterm exam      & 48.3 & 36.0 & 12.3 & [-4.2, 28.5] \\
Second midterm exam      & 41.7 & 35.5 &  6.2 & [-10.3, 22.9] \\
Global exam     & 55.6 & 32.6 & 22.9 & [5.1, 39.5] \\
Final mark & 67.7 & 45.1 & 22.6 & [6.6, 37.2] \\
\bottomrule
\end{tabular} 
\caption{Pooled student-level pass rates and risk difference (AL$-$TL) presented in percentages.}
\label{tab:pooled-summary}
\end{table}

Results in Figure~\ref{fig:forest} and Table~\ref{tab:pooled-summary} indicate that, for instance, the pooled risk difference for the final mark is about $23$ percentage points (AL$-$TL), meaning that roughly 23 \emph{additional} students (or as few as 7 and as many as 37 additional students in a 95\% confident interval) pass the course per 100 under AL compared with TL under this pooled benchmark. It must be emphasized that since the final mark includes a substantial instructor-determined component, differences in course pass rates may reflect both performance on shared assessments and differences in section-level grading/activities; we therefore recommend in this case to treat the global exam as the cleanest common-assessment benchmark and the final mark as the most policy-relevant but less standardized outcome.

This analysis thus far uses aggregated counts and a working  binomial/independence approximation. As a consequence, it provides a transparent and reproducible summary of modality differences in pooled pass rates, but it does not attempt to separate modality effects from other sources of variation that are not identifiable from totals alone. In particular, if student outcomes are correlated within sections (clustering) or if modality is confounded with instructor or scheduling effects or heterogeneous rubrics, the nominal standard errors from an independence model may be optimistic. Addressing that requires either student-level data or explicit cluster-aware modeling (e.g., cluster-robust inference, hierarchical models) along with enough sections per modality to estimate between-section variability credibly. Within these limits, the outputs should be interpreted as follows: the pooled pass-rate bars estimate the fraction of students who passed under each modality for each assessment; the Wilson error bars quantify binomial sampling uncertainty for those pooled proportions; and the forest plot (Figure~\ref{fig:forest}) expresses the modality effect as an absolute difference in pass probability, with uncertainty quantified for the difference itself.

We next present a Bayesian re-expression of the same pooled estimands to provide posterior uncertainty summaries on the probability scale and to connect naturally to the hierarchical sensitivity analysis. The resulting Bayesian summaries are reported in Table~\ref{tab:bayes-jeffreys} and are consistent with the previous frequentist approach, with the advantage that it naturally contains all the statistical information in the posterior distribution. Uncertainty summaries are obtained directly from Monte Carlo draws of the conjugate posteriors.

\begin{table}[h!]
\centering
\small
\setlength{\tabcolsep}{3.5pt}
\begin{tabular}{@{}lccccccc@{}}
\toprule
Assessment &  AL pass & 95\% CrI & TL pass & 95\% CrI & RD & 95\% CrI & $\text{Prob}(\mathrm{RD}>0)$ \\
\midrule
First midterm  & 48.3 & [38.1, 58.6] & 36.1 & [29.9, 42.4] & 12.3 & [0.3, 24.3]  & 97.74\% \\
Second midterm & 41.8 & [31.6, 52.3] & 35.6 & [29.3, 42.0] &  6.2 & [-5.8, 18.5] & 84.06\% \\
Global exam    & 55.5 & [44.7, 66.0] & 32.7 & [26.3, 39.5] & 22.8 & [10.1, 35.2] & 99.98\% \\
Final mark     & 67.5 & [57.9, 76.4] & 45.1 & [39.0, 51.2] & 22.5 & [11.1, 33.3] & 99.99\% \\
\bottomrule
\end{tabular}
\caption{Bayesian pooled pass-rate analysis with Jeffreys prior $\mathrm{Beta}(\tfrac12,\tfrac12)$. Intervals are central 95\% posterior credible intervals.}
\label{tab:bayes-jeffreys}
\end{table}

Table~\ref{tab:hierarchical_passrate} reports the resulting hierarchical  pooled pass-rate estimates. Posterior summaries under the hierarchical model are highly consistent with the pooled frequentist results reported above. For the available data, accounting for section-level clustering  does not seem to materially change the estimated pooled student-weighted pass probabilities $p_{\beta,a}$ or the corresponding risk differences $\mathrm{RD}_a$. This agreement is expected because the hierarchical model targets the same pooled estimands via Eq.~\eqref{eq:hier_pool}, and the inferred level of section-to-section variability is not large enough to shift the pooled means away from the aggregated rates; rather, it mainly affects the width of the posterior intervals. We therefore view the hierarchical analysis as a robustness check: the qualitative conclusion that AL sections exhibit higher pass rates across assessments remains unchanged when allowing for within-section dependence. More generally, the section random intercept $u_\ell$ is a catch-all for persistent section-to-section differences, and with only aggregated section-level pass counts it cannot be decomposed into specific mechanisms. Access to richer covariates---for example, instructor experience, time-of-day, section size, prior academic preparation (e.g., placement/diagnostic scores), GPA proxies, repeat status, attendance/engagement measures, or assignment/quiz performance---would allow clustering and heterogeneity to be modeled and interpreted more directly (and, in some cases, partially explained rather than absorbed into $u_\ell$). Collecting such covariates in future trimesters would therefore enable more informative cluster-aware analyses, including estimating how much of the observed between-section variability is attributable to measurable factors versus residual idiosyncratic section effects.

\begin{table}[t]
\centering
\begin{tabular}{lccccccc}
\hline
Assessment & AL pass &95\% CrI  & TL pass  &95\% CrI  & RD & 95\% CrI (pp) & $\text{Prob}(\mathrm{RD}>0)$ \\
\hline
1st midterm   & 48.2 & [35.7, 60.6] & 36.1 & [30.7, 42.0] & 12.1 & [0.4, 23.6]  & 97.92\% \\
2nd midterm  & 41.6 & [29.5, 54.7] & 35.6 & [29.9, 41.6] & 6.0  & [-5.3, 18.0] & 84.17\% \\
Global          & 55.3 & [41.9, 68.0] & 32.8 & [27.0, 39.0] & 22.5 & [10.4, 34.3] & 99.98\% \\
Final mark   & 67.6 & [55.6, 78.2] & 45.2 & [39.4, 51.1] & 22.4 & [11.2, 32.6] & 99.99\% \\
\hline
\end{tabular}
\caption{Bayesian hierarchical (section-clustered) pass-rate analysis. Reported intervals are central 95\% posterior credible intervals. The risk difference is $\mathrm{RD}=p_{\mathrm{AL}}-p_{\mathrm{TL}}$ (percentage points), and $\text{Prob}(\mathrm{RD}>0\mid \text{data})$ is estimated by Monte Carlo draws from the fitted variational Bayes approximate posterior.}
\label{tab:hierarchical_passrate}
\end{table}

%%%%%%%%%%%%%%%%%%%%%%%%%%%%%%%%%%%%%%%%%%%%%%%%%%%%%%%%%%%%%%%%%%%%%%%%%%%%%%
\section{Conclusions and Outlook}
\label{sec:conclusion}
%%%%%%%%%%%%%%%%%%%%%%%%%%%%%%%%%%%%%%%%%%%%%%%%%%%%%%%%%%%%%%%%%%%%%%%%%%%%%%
This paper compares passing outcomes between sections implemented with Active Learning (AL) and sections implemented with Traditional Lecturing (TL) in \textit{Elementary Mechanics I} during a single trimester, with an unusual student population in PER, under common departmental assessments. Using only aggregated counts from official section reports, we estimated pooled pass probabilities for each modality across the two midterm exams, the common global exam, and the final mark, and summarized modality differences primarily on an absolute scale via the risk difference \(RD_a=p_{\mathrm{AL},a}-p_{\mathrm{TL},a}\), complemented by uncertainty quantification using both Wilson confidence intervals and a Bayesian reference analysis with Jeffreys priors. Across assessments, the pooled estimates consistently favored AL-labeled sections, with the strongest separation observed on the common global exam and in the final mark under the shared assessment standard.

The main contribution is not a claim of a universal effect size, but a documented institutional implementation and a reproducible outcome benchmark of ``AL as implemented'' versus ``TL as implemented'' in this setting. The results are encouraging in light of the practical constraints of early adoption: the instructors leading AL implementation faced a steep learning curve, and the AL sections were supported by additional preparation time and TA staffing that are relevant for realistic scaling. At the same time, the data structure imposes a clear interpretation boundary. Because the design is not randomized and the available information is aggregated at the section level, the analysis does not identify the causal effect of any single instructional technique and cannot adjust for student self-selection into sections, section-level clustering, or instructor and scheduling effects. The appropriate interpretation is therefore descriptive and decision-oriented: within term 25-O and under common departmental assessments, students enrolled in AL-labeled sections exhibited higher pooled passing probabilities than students enrolled in TL-labeled sections, with uncertainty quantified under a working binomial model.

From the standpoint of departmental decision-making, these results support three immediate implications. First, when outcomes in gateway courses are a central institutional concern, a supported AL implementation can be associated with higher passing outcomes even in an early adoption phase. Second, the common-assessment structure is a practical asset: it enables cross-section comparison without relying on instructor-specific grading practices, and it provides a feasible reporting pipeline that can be repeated term-to-term. Third, the initiative highlights that implementation supports matter for sustainability: instructor preparation, coordinated materials, and TA staffing were integral components of what ``AL'' meant in practice and should be treated as part of the intervention rather than as incidental details. The impacts of particular elements of this change model are left for a future qualitative study of the initiative. 

A natural next step is to strengthen evaluation while keeping data collection feasible. Minimal upgrades that would substantially improve interpretability include: (i) routine COPUS-based documentation of classroom practice across a larger sample of meetings and sections to provide a clearer fidelity record; (ii) collection of a small number of student-level covariates available in administrative systems (e.g., prior course attempts, placement indicators, or program/major) to assess composition differences across sections; and (iii) adoption of a simple cluster-aware analysis plan (e.g., section-level resampling or hierarchical modeling) once enough AL sections are available to estimate between-section variability credibly. In parallel, future work should move beyond pass/fail outcomes by incorporating concept inventories or common item-level measures where feasible, to connect observed passing outcomes more directly to conceptual learning.

Similarly, quasi-experimental quantitative analyses such as this may be necessary but not sufficient to judge the efficacy of an institutional change initiative--particularly because qualitative data is often more convincing for individual instructors to change habits. We are currently conducting a qualitative study of the initial change process with the local change agents--faculty and TAs involved in the initiative--using Reinholz's ``four frames'' applied to educational change \cite{fourframes-2018, https://doi.org/10.1002/sce.21526}. This will complement the quantitative analyses to help identify institutional structures which may support or hinder the change initiative as it moves forward and expands to other departments.

In summary, within the constraints of aggregated institutional reporting, the term 25-O results provide a clear benchmark suggesting that active-learning implementation, supported by structured preparation and staffing, can be associated with higher pooled passing outcomes in introductory mechanics under common departmental assessments. The combination of an operational definition of AL/TL, transparent pooled estimands, and explicit interpretation boundaries offers a practical template for departments seeking to evaluate instructional change while building toward more rigorous designs in subsequent terms.

\acknowledgments
I.P.C. thanks Peter LePage and Carl Wieman for guidance in developing the initiative workplan, drafting support letters, and facilitating introductions to Eric Burkholder and Louis Deslauriers. He also thanks Louis Deslauriers for ongoing guidance on the active-learning initiative and for leading the second two-day workshop for instructor and TA training. The three workshops were supported by local funding from the Programa Especial de Apoyo a Proyectos de Docencia e Investigación (PEAPDI) 2025. We also thank the Coordinación Divisional de Docencia y Atención al Alumnado de la División de Ciencias Básicas e Ingeniería for supporting this initiative.

\bibliographystyle{apsrev4-2}  
\bibliography{bio}

\end{document}